\documentclass[prb,amsmath,amssymb,reprint,twocolumn,superscriptaddress,showpacs]{revtex4-2}
\usepackage{graphicx}
\usepackage{dcolumn}
\usepackage{verbatim}
\usepackage{bm}
\usepackage[
bookmarks=true,
colorlinks,
linkcolor=blue,
urlcolor=blue,
citecolor=blue,
plainpages=false,
pdfpagelabels,
final,
breaklinks=true
]{hyperref}

\begin{document}

\preprint{APS/123-QED}

\title{Geometric-anisotropy induced high-order topological insulators in nonsymmorphic photonic crystals}

\author{Zhenzhen Liu}
\author{Guochao Wei} 
\author{Jun-Jun Xiao}%
 \email{eiexiao@hit.edu.cn}
\affiliation{Shenzhen Engineering Laboratory of Aerospace Detection and Imaging, College of Electronic and Information Engineering, Harbin Institute of Technology (Shenzhen), Shenzhen 518055, China
}

\date{\today}
\begin{abstract}
To a significant extent, the rich physical properties of photonic crystals are determined by the underlying geometry, in which the composed symmetry operator and their combinations contribute to their unique topological invariant to characterize the topological phases. Particularly, the inter- and intra-coupling modulation in the two-dimensional (2D) Su-Schrieffer-Heeger model yields the topological phase transition, and exhibit first-order edge localized states and second-order corner localized corner states. In this work, we use the geometric anisotropy into the 2D square lattice composed of four rectangle blocks. We show a variety of topological phase transitions in designed nonsymmorphic photonic crystals (PCs) and these transitions shall be understood in terms of the Zak phase and Chern number in synthetic space, as well as the pseudospin-2 concept, combinationally. Furthermore, Zak phase winding in the periodic synthetic parameter space yields high-order Chern number and double interface states. Based on the extended Zak phase and pseudo-spin Hall effect, higher-order topological insulator is constructed in the PC system. The intriguing and abundant topological features are also sustained in the corresponding three-dimensional PC slab, which makes it a very interesting platform to control the flow of optical signals.
\end{abstract}

\maketitle

%\newpage
\section{Introduction}

Topological physics has attracted great attention in the past decades~\cite{Zak1989Berrys, Hasan2010Colloquium, Qi2008Topological, Qi2011Topological, Xiao2010Berry, Benalcazar2017Quantized}. The concept of nontrivial topological phase has been extended to classical systems, such as topological photonic, acoustic, and elastic waves systems, with particular interests in topological edge states~\cite{Haldane2008Possible,Wang2009Observation, Wu2015Scheme,Proctor2019Exciting, Yang2018Visualization,Deng2017Observation, Wang2020Effective, Fan2020Elastic,Yu2018Elastic}. 
Chiral edge states have also been predicted for electromagnetic waves \cite{Haldane2008Possible} and experimentally confirmed using gyromagnetic materials under an external magnetic field~\cite{Wang2009Observation, Skirlo2015Experimental}. Alternatively, analogues for the quantum spin Hall effect (SHE) and valley Hall effect that do not require breaking of the time-reversal symmetry, have been achieved by appropriately controlling the spatial symmetry~\cite{Wu2015Scheme, Proctor2019Exciting, Xia2017Topological,Dong2016Valley, Wu2017Direct}.

In two-dimensional (2D) photonic crystals (PCs), the SHE is the result of the topological phase transition accompanied with the interchange of two doubly-degenerate points at high-symmetry momenta~\cite{Wu2015Scheme, Huang2019Reconfigurable}. Generally, the pseudo-time reversal symmetry satisfying the Kramer's double degeneracy ensures the double degeneracy~\cite{Wu2015Scheme, Wang2017Type, Tsirkin2017Composite}. For example, triangle lattice obeying $C_6$ symmetry yields Dirac cones at the Brillouin zone (BZ) corners. In the honeycomb lattice, these Dirac cones are projected to the BZ center which gives rise to a 4-fold degeneracy and yields the SHE~\cite{Wu2015Scheme}. Band inversion by expanding/shrinking~\cite{Barik2016Two, Proctor2019Exciting} or scaling~\cite{Deng2017Observation, Ji2019Transport} the meta-atoms could lead to topological phase transition between the trivial insulator and the quantum spin Hall insulator, describable by the spin Chern number~\cite{Wu2015Scheme}. 

Beyond the traditional bulk-edge correspondence, higher-order topological insulator has been found which hosts lower-dimensional boundary states \cite{Benalcazar2017Quantized, Benalcazar2017Electric}. For example, second-order topological insulator has gapped one-dimensional (1D) edge states and zero-dimensional (0D) corner states in the gap \cite{Xie2018Second, Zhang2019Second, Chen2019Direct}. High-order topological insulators broaden the scope of topological material and enrich the understanding of band topology. Many theoretical and experimental works have been done on high-order topological insulators. In terms of the highly localized topological corner state, a nano-cavity with extremely high Q-factor \cite{Ota2019Photonic} and topological laser with low-threshold \cite{Zhang2020Low} have been designed and experimentally realized. 

\begin{figure}[t]
\includegraphics[width=8.5cm]{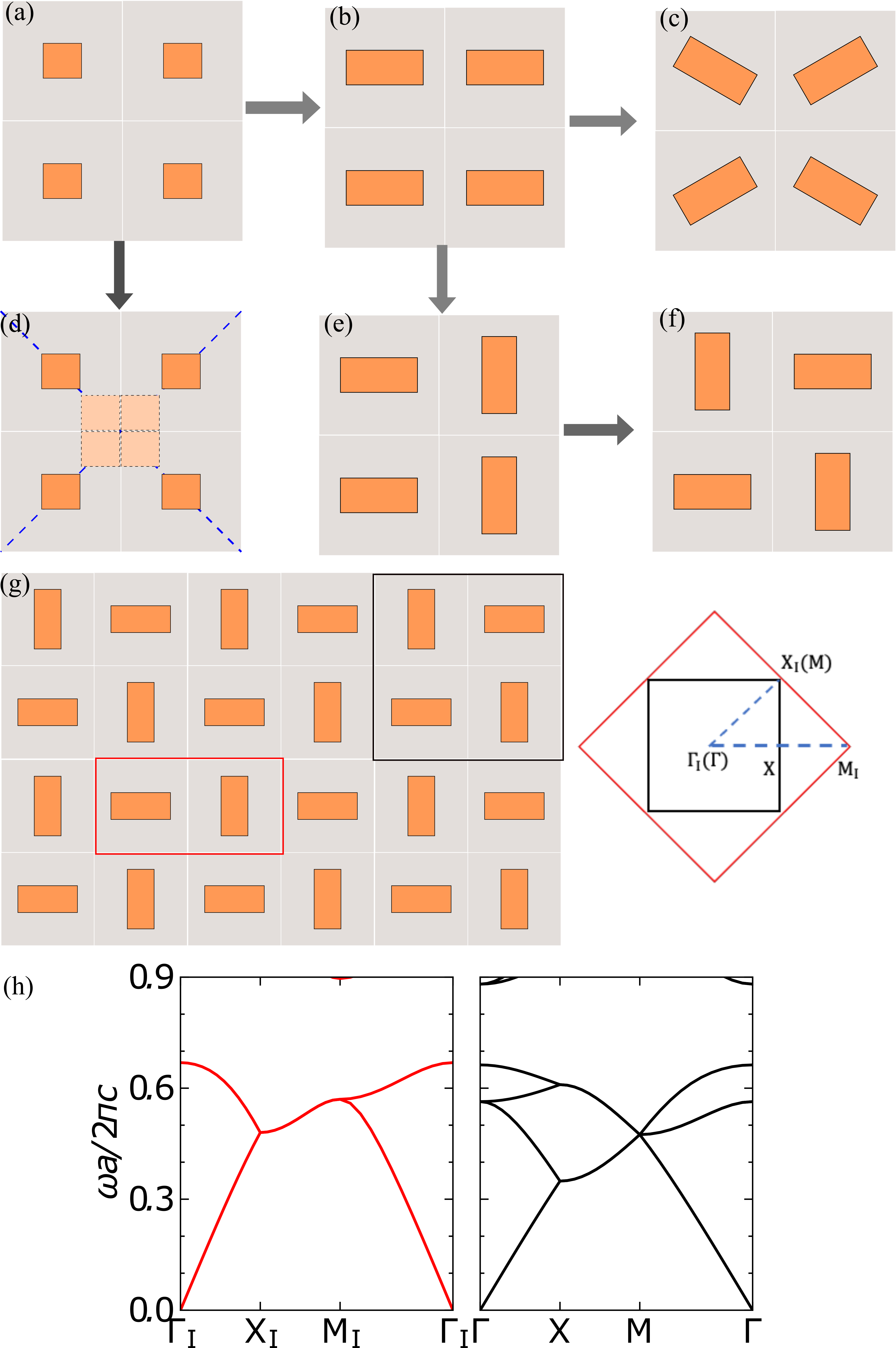}
\caption{\label{fig:1} Geometric transition scheme in 2D square lattice PCs consisting of rectangular meta-atoms with permittivity $\varepsilon_r$. The original square blocks (a) in each unit cell can be transformed into rectangular blocks (b) or move away from unit center (d). For rectangular block, rotating the adjacent four blocks (c) enlarges the unit cell in both directions and rotating one block (e) only enlarges the unit cell in one specific ($x$) direction. Furthermore, the rotation of the diagonal blocks (f) introduces the geometric anisotropy, whose periodic cell can be taken as a primitive unit cell [red box in (g)] and an enlarged supercell [black box in (g)], whose BZ are, respectively, shown in the right panel of (g). (h) The corresponding band structure of the primitive unit cell and the enlarged one. High symmetry ${\bf k}$-points are defined in the corresponding BZ.}
\end{figure}

Different mechanisms of the high-order topological insulators have been revealed, mostly featured by either quantized multipole moments \cite{Benalcazar2017Quantized, Benalcazar2017Electric, He2020Quadrupole} or generalized Su-Schrieffer-Heeger (SSH) models without quantized multipole moment \cite{Hassan2019Corner, Ota2019Photonic}. For 2D square photonic crystals describable by generalized SSH model, the topological invariant Zak phase $(Z_x,Z_y)$, can be tuned by moving the constituent blocks within four adjacent unit cells away from their centers, as shown in Figs.~\ref{fig:1}(a) and \ref{fig:1}(d). Besides, the quadrupole moment insulator can be constructed by breaking time-reversal symmetry with substituted gyromagnetic rods~\cite{He2020Quadrupole}. From the perspective of multidimensional topological phases, topology of their topological boundary states yields the lower-dimensional corner states. For instance, the gapped helical edge states from SHE carry quantized Zak phases, which leads to zero-dimensional topological corner states~\cite{Zhang2019Second}. This configurable geometry can be obtained by rotating the transformed rectangular blocks, as shown in Figs.~\ref{fig:1}(b) and \ref{fig:1}(c). However, mirror symmetry plays a significant role in the characteristics of topological invariant, which limits the geometrical configuration. There exist many other schemes to configure the geometry, as sketched in Fig.~\ref{fig:1}(f). 

Considering the usefulness of the band folding and glide symmetry, in this work, we propose an alternative way to explore the SHE and the resultant second order insulator. Interestingly, we find that different choices of the unit cell exhibit distinct topological properties characterized by the bulk polarization, as well as multidimensional topological physics. The proposed PCs could provide a promising platform for the development of versatile conventional and high-order topological photonic applications. We prove the possibility of transferring the results to PC slab systems that are more attractive for on-chip application.

\section{Conventional Topological Features of PC}

Figures~\ref{fig:1}(a)-\ref{fig:1}(f) schematically show several typical geometrical transformation of a square lattice composed of isolated rectangular meta-atoms. For geometrically anisotropic structures, the schematic of square-lattice PCs is shown in Fig.~\ref{fig:1}(g). Here, the primitive unit cell marked by the red box hosts two meta-atoms, whose BZ is denoted in the right panel with high symmetry momentum line along $\Gamma_1-\mathrm{X}_1-\mathrm{M}_1$. Note that the bands along the boundary $\mathrm{X}_1\mathrm{M}_1$ are doubly degenerate due to the combination of glide symmetry and time-reversal symmetry \cite{Xia2019Observation}. Whereas, the enlarged unit cell framed by the black box possesses four meta-atoms. In this case, the BZ becomes shrunken, which can be treated as the zone folding of the original BZ along the black lines detailed in Appendix \ref{appendix:fold}. In this regard, four-fold degeneracy at M and double degeneracy along XM are constructed, as illustrated in the band structure shown in Fig.~\ref{fig:1}(h). We note that this establishes the starting point of the following analysis. 

\begin{figure}[t]
\includegraphics[scale=0.9]{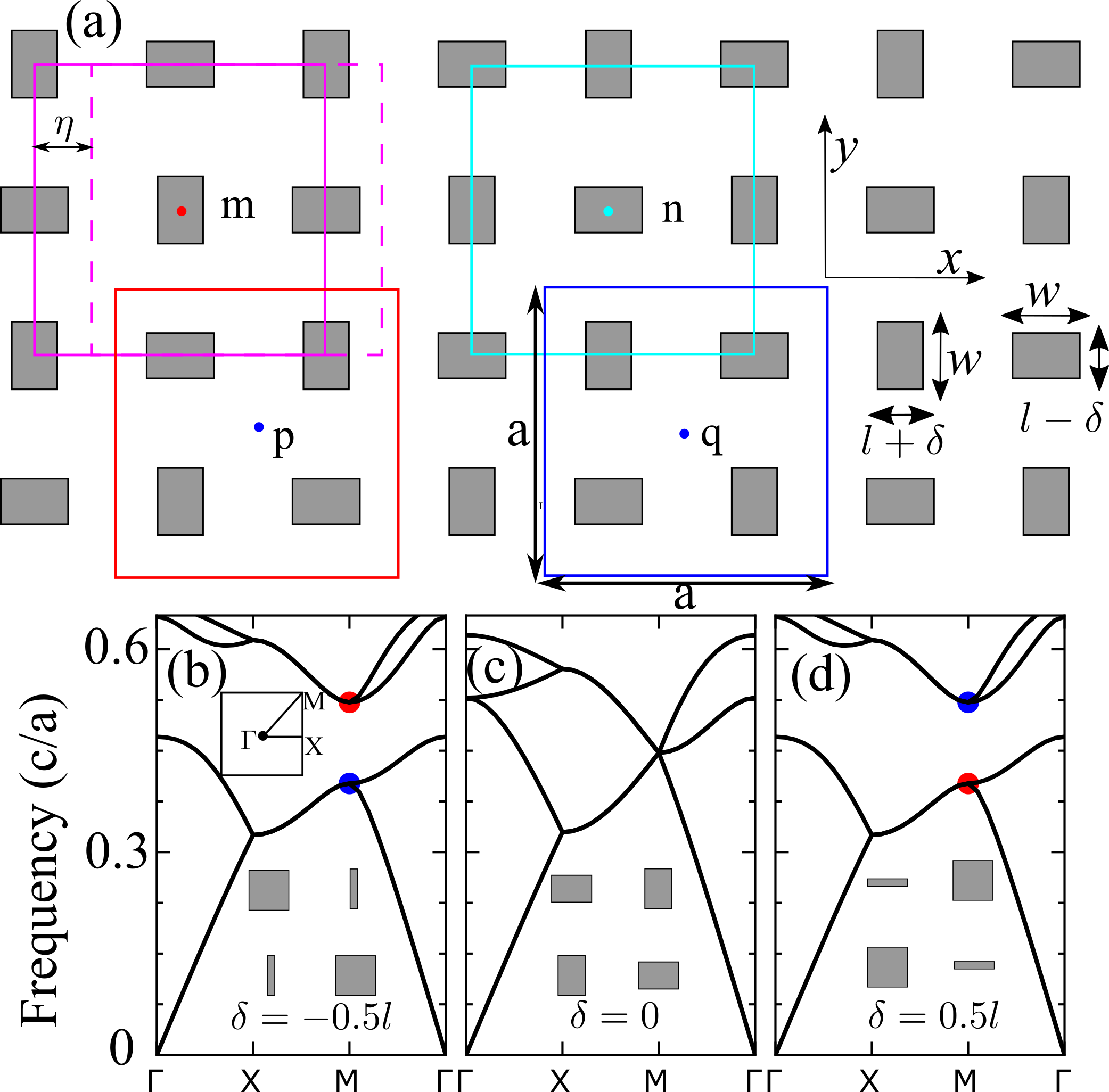}
\caption{\label{fig:fig2} Schematic of the square lattice PCs consisting of rectangular meta-atoms with permittivity $\varepsilon_r$ and the evolution of the bandgap upon geometrical variation. (a) Four inequivalent supercells (colored box) centered at the rectangle block (labeled as $m$ and $n$) or the interspace (labeled as $p$ and $q$). Photonic band structure of the TM mode with $\varepsilon_r=9.5$, $l=0.15a$, and $w=0.225a$ for (b) $\delta=-0.5l$ , (c) $\delta=0$ , and (d) $\delta=0.5a$. Here $\delta$ is a geometry tuning parameter and the insets in (b)-(d) are the $p$-type supercell with three typical $\delta$.}
\end{figure}

Figure~\ref{fig:fig2}(a) shows the schematic diagram of the 2D PC consisting of rectangular meta-atoms whose long axis orientates alternatively toward $x$ and $y$ directions. The lattice constant is $a$, and the width and height of the rectangle blocks are, respectively, $w$ and $l$. As the PC is invariant under the interchange of $x$- and $y$-oriented rectangle blocks, its structure is self-complementary~\cite{Chen2018Edge}. There are four preliminary ways to construct the PC by choosing different unit cell. One type is of unit cell positioned at the center of the rectangle block, labeled as $m$ and $n$ [outlined respectively by the purple and cyan squares in Fig.~\ref{fig:fig2}(a)], whose meta-atoms are with mirror symmetry. The other type is with unit cell centered at the air (red and blue squares marked by point $p$ and $q$), whose meta-atoms are without mirror symmetry. The geometry parameter $\delta$ acting on the sizes of the blocks (e.g., $l\to l+\delta$ for $y$-oriented blocks and $l\to l-\delta$ for $x$-oriented blocks) is tuned to investigate the band evolution, and the continuous transnational displacement $\eta$ makes connections between different unit cells [see the red dashed and solid squares in Fig.~\ref{fig:fig2}(a)]. As such, we denote the various PCs formed by periodically repeating the $m$, $n$, $p$, $q$ centered lattice as $\mathrm{PC}_{\delta,\eta}^{(m,n,p,q)}$.

\begin{figure}[t]
\includegraphics[scale=0.9]{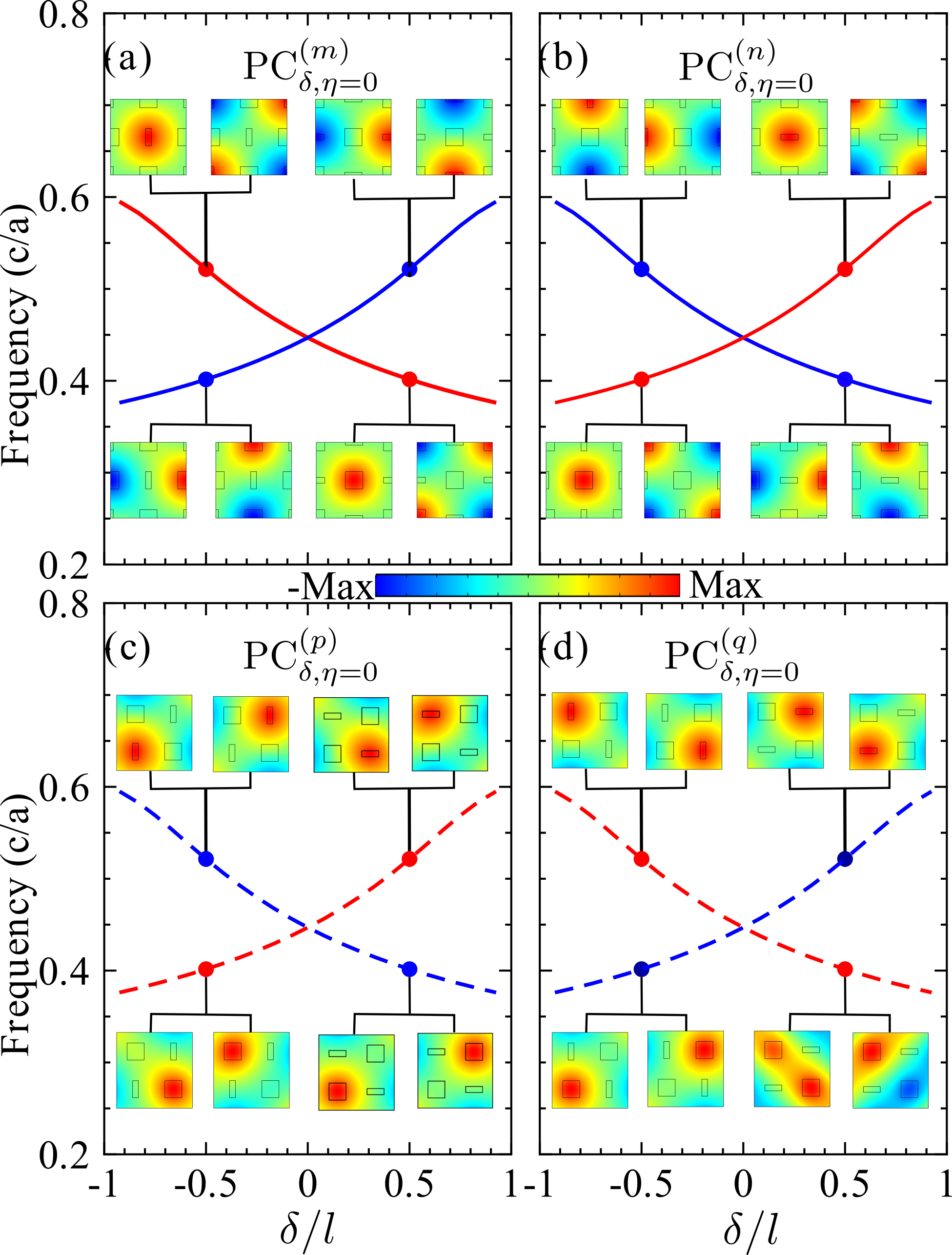}% Here is how to import EPS art
\caption{\label{fig:fig3} Topological phase diagram for (a) $\mathrm{PC}_{\delta,0}^{(m)}$, (b) $\mathrm{PC}_{\delta,0}^{(n)}$, (c) $\mathrm{PC}_{\delta,0}^{(p)}$ and (d) $\mathrm{PC}_{\delta,0}^{(q)}$ at momentum M for varying $\delta$. The frequency of the odd-parity doublet and that of the even-parity doublet are plotted as red and blue curves, respectively. Here, the terms ‘even’ and ‘odd’ indicate the eigenvalues being +1 and -1 for corresponding symmetry operations on the eigenstates, respectively, under (a) and (b) inversion operator, (c) and (d) rotation-translation operator. The inset mode profiles Re($E_z$) show the eigenmodes at points marked by the red and blue dots. }
\end{figure}

Here, the pseudo-time reversal symmetry operator $G_xT$ satisfies $(G_xT)^2=-1$, which guarantees the appearance of Kramers doublet at M point (details in Appendix \ref{appendix:fold}). For $\delta\ne 0$, the glide symmetry of the primitive unit cell is broken, while the glide symmetries of the enlarged unit cell are preserved. Consequently, the 4-fold degeneracy is lifted and double degeneracy remains along the BZ boundary XM. Results are shown in Figs.~\ref{fig:fig2}(b)-\ref{fig:fig2}(d) for $\delta=-0.5l$, $\delta=0$, and $\delta=0.5l$, respectively. The topological phase diagrams of $\mathrm{PC}_{\delta,0}^{(m,n,p,q)}$ are illustrated in Figs.~\ref{fig:fig3}(a)-\ref{fig:fig3}(d), respectively. Due to the commutation relation between the inversion operator and the glide operator $G_{x}:= (x,y)\rightarrow(-x+a/2,y+a/2)$) and $G_{y}:= (x,y)\rightarrow(-y+a/2,x+a/2)$ for $\mathrm{PC}_{\delta,0}^{(m,n)}$, the parity of the eigenmodes at the point $\text{M}$ can be featured by the eigenvalue of the inversion operator~\cite{Zhang2019Second}. The red and blue dots in Figs.~\ref{fig:fig3}(a) and~\ref{fig:fig3}(b) represent the odd-parity (with eigenvalue -1) and even-parity (with eigenvalue +1), respectively. The odd-parity eigenstates consist of the $p_x$ and $p_y$ like states, and the even-parity eigenstates consist of the $s$ and $d$ like states, as shown in the insets of Figs.~\ref{fig:fig3}(a) and~\ref{fig:fig3}(b). With the exchange of the even- and odd-parity doublet, two distinct phases emerge during the band crossing. From the perspective of SHE, the bandgap is with trivial topological phase when the frequency of the odd-parity eigenstates is larger than that of the even-parity eigenstates~\cite{Zhang2019Second}. On the contrary, it is with topological phase. Clearly, Fig.~\ref{fig:fig3}(a) shows the phase transition from topological phase to topological trivial phase in term of the SHE. In contrast to $\mathrm{PC}_{\delta,0}^{(m)}$, $\mathrm{PC}_{\delta,0}^{(n)}$ exhibits the opposite case of transition from trivial phase to nontrivial phase, as $\delta$ increases. This gap closing and reopening for the two pairs of degeneracy is a reminiscent of the photonic SHE~\cite{Wu2015Scheme}. 

However, the situation is quite different for $\mathrm{PC}_{\delta,0}^{(p)}$ and $\mathrm{PC}_{\delta,0}^{(q)}$. Instead of the predefined operator $G_x$, the transformed PCs obey an alternative glide symmetry $G^{'}_x:=(x,y)\rightarrow (-x,y+a/2)$ and $G^{'}_y:=(x,y)\rightarrow (x+a/2,-y)$. Since the eigenvalue of $G^{'}_y$ along $k_x=\pi/a$ is $\pm i$ and the inversion operator anti-commutates with the operator $G'_y$, so the eienstates of the glide operator $G^{'}_y$ is not compatible with the inversion operator \cite{Takahashi2017Spinless}. Alternatively, a new rotation-translation operator $S_2:=(x,y)\rightarrow (a/2-x, a/2-y)$ is applied to elucidate the analogous topological phase transition, since this operator commutates with the glide operator $G^{'}_y$. As a result, the eigenstates of the glide operator can be conceptually described by the eigenvalue of the rotation-translation operator $S_2$, i.e., $+1$ for even parity (red dashed curve) and $-1$ for odd-parity (blue dashed curve) in Fig.~\ref{fig:fig3}(c) which shows the corresponding topological phase diagram at point $\text{M}$, similar to Fig.~\ref{fig:fig3}(b). Note that the even- and odd-parity cannot be directly distinguished by the eigen-fields in the inset of Fig.~\ref{fig:fig3}(c). Opposite to $\mathrm{PC}_{\delta,0}^{(p)}$, $\mathrm{PC}_{\delta,0}^{(q)}$ exhibits a topological phase transition from nontrivial phase to trivial phase, as shown in Fig.~\ref{fig:fig3}(d).

\begin{figure}[t]
\includegraphics[scale=0.9]{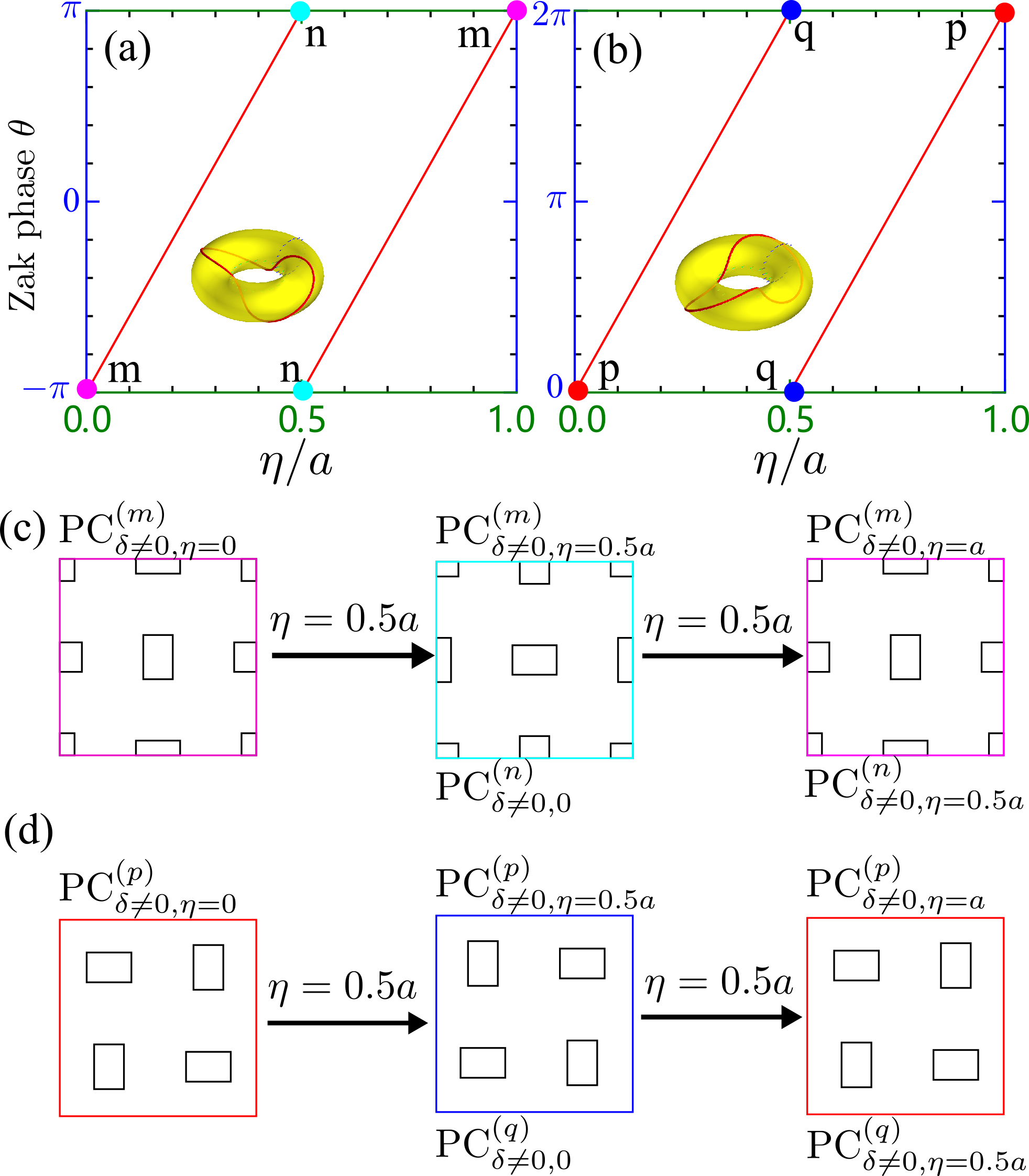}% Here is how to import EPS art
\caption{\label{fig:fig4} Zak phase evolution as a function of $\eta$ for (a) $\mathrm{PC}_{\delta,\eta}^{(m)}$ and (b) $\mathrm{PC}_{\delta,\eta}^{(p)}$. Illustration of the relation between the transnational unit cell with $\eta=0.5a$ for (c) $\mathrm{PC}_{\delta,\eta}^{(m)}$ and (d) $\mathrm{PC}_{\delta,\eta}^{(p)}$. The torus in (a) and (b) are constructed by gluing the equivalent edges for the two gauge dimensions $k_x$ and $\eta$.}
\end{figure}

To this end, topological phase transition and the resultant SHE have been demonstrated in $m$,$n$,$p$,$q$-centered PCs and illustrated by the band inversion with varying $\delta$. However, for arbitrary unit cell selection, particularly when $\eta \neq 0$ or $\eta \neq a/2$ , the band inversion is not readily apparent. As a matter of fact, during the continuous spatial translation by $\eta$, the lattice geometry is no longer inversion-symmetric, therefore the Zak phase varies from $-\pi$ to $\pi$~\cite{Nakata2020Topological, Lu2021Topological}. Let us first consider $\mathrm{PC}_{\delta,\eta}^{(m)}$. Notice that the displacement parameter $\eta$ is periodic and can be regarded as a synthetic dimension~\cite{Nakata2020Topological,Lu2021Topological, lu2022chip}. Combined with the Bloch wave vector $\left( k_x,k_y\right)$, three-dimensional parameter space $\left(k_x,k_y,\eta\right)$ is formed. If the time-reversal symmetry is broken, non-trivial Chern number can be quantized in the momentum space~\cite{Skirlo2015Experimental}. For an arbitrary $k_y$, the 2D parameter space $(k_x,\eta)$ forms a torus as shown in the inset of Figs.~\ref{fig:fig4}(a) and~\ref{fig:fig4}(b). The non-Abelian Zak phase for the first band gap is calculated via \cite{Resta2000Manifestations} 
\begin{equation}
\label{eq:eq1}
\theta(k_y,\eta)=\int_{-\frac{\pi }{a}}^{\frac{\pi }{a}}\textbf{Tr}\langle u_m(k_x,k_y,\eta)\vert \frac{i\partial }{\partial {k_{y}}}\vert u_n(k_x,k_y,\eta)\rangle dk_x,
\end{equation}
where $\vert u_n({k_x,k_y,\eta})$ is the periodic part of the Bloch state of the \textit{n}-th band.

The phase winding of $\theta(k_y,\eta)$ defined in Eq.~(\ref{eq:eq1}) counts the Chern number
\begin{equation}
C({k_y})=\frac{1}{2\pi }\int_{0}^{a}{{{\partial }_{\eta }}}\theta ({k_y},\eta )d\eta,
\end{equation}
which represents the topological characteristics on the $(k_x,\eta)$ plane \cite{Asboth2016Short}. Figure~\ref{fig:fig4}(a) shows the trajectory of the Zak phase $\theta(k_y,\eta)$ of $\mathrm{PC}_{\delta,\eta}^{(m)}$ with $\delta=-0.5l$ and $k_y=0$, as $\eta$ goes from $\eta=0$ to $\eta=a$. Clearly, the total Zak phase changes by $4\pi$ for this variation and yields Chern number $C(k_y)=2$. More importantly, this holds for all fixed $k_y$. Similarly, the phase winding for $\mathrm{PC}_{\delta,\eta}^{(p)}$ is $4\pi$ as shown in Fig.~\ref{fig:fig4}(b). The difference is that the Zak phase $\theta=\pm \pi$ when $\eta=0$ for $\mathrm{PC}_{\delta,0}^{(m)}$, while for $\mathrm{PC}_{\delta,0}^{(p)}$ the Zak phase $\theta=0$ or $2\pi$ when $\eta=0$. According to the bulk-edge correspondence, a pair of $\eta$-dependent interface state can be constructed as $\eta$ varies. Appendix \ref{appendix:chern} exemplifies the presence of interfaces states determined by synthetic Chern number.

In both Figs.~\ref{fig:fig4}(a) and~\ref{fig:fig4}(b) there exists a special transition point at $\eta=0.5a$, where $\theta(k_y,0.5a)=\pi$ for $\mathrm{PC}_{\delta,0.5a}^{(m)}$ and $\theta(k_y,0.5a)=0$ for $\mathrm{PC}_{\delta,0.5a}^{(p)}$. As illustrated in Fig.~\ref{fig:fig4}(c), $\mathrm{PC}_{\delta,0}^{(n)}$ is actually identical to $\mathrm{PC}_{\delta,0.5a}^{(m)}$. Similarly, Fig.~\ref{fig:fig4}(d) shows that $\mathrm{PC}_{\delta,0}^{(q)}$ is exactly $\mathrm{PC}_{\delta,0.5a}^{(p)}$. This suggests that one deterministic interface state is present between PCs constructed with $m$ or $n$-centered unit cell and $p$ or $q$-centered unit cell due to the different Zak phase~\cite{Huang2014Sufficient,Chen2018Edge}. 

With respect to the geometrical tuning parameter $\delta$, spatial displacement $\eta$, and the four special unit cell selections with and without mirror symmetry, the topological invariants are therefore described by the pseudo-spin phase transition, Zak phase, and Zak phase winding in synthetic space (or Chern number). When the band gap above the two lowest bulk bands is considered, deterministic interface states can be found at the boundary between these PCs with different topological origin. 

\begin{figure}[t]
\includegraphics[scale=0.9]{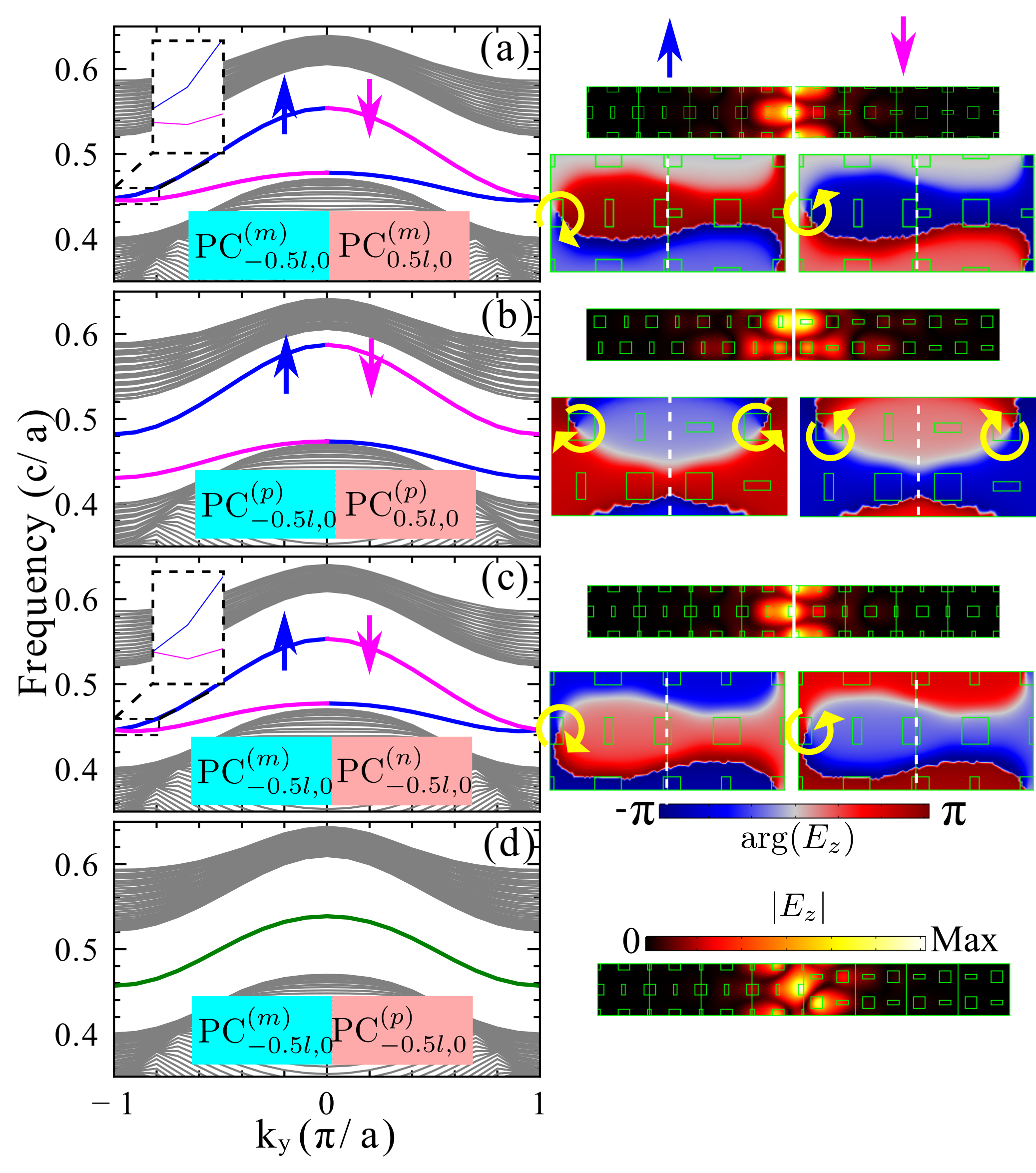}% Here is how to import EPS art
\caption{\label{fig:fig5} Projected band diagram for the interfacial structures shown in the inset constructed by combining (a) $\mathrm{PC}_{-0.5l,0}^{(m)}$ and $\mathrm{PC}_{0.5l,0}^{(m)}$, (b) $\mathrm{PC}_{-0.5l,0}^{(p)}$ and $\mathrm{PC}_{0.5l,0}^{(p)}$, (c) $\mathrm{PC}_{-0.5l,0}^{(m)}$ and $\mathrm{PC}_{-0.5l,0}^{(n)}$ and (d) $\mathrm{PC}_{-0.5l,0}^{(m)}$ and $\mathrm{PC}_{-0.5l,0}^{(p)}$. Amplitude $\vert E_z \vert$ and phase $\mathrm{arg}(E_z)$ distribution are for the interface states marked by the arrows in (a-d). Periodic (absorbing) boundary condition is applied on the $y$ ( $x$) direction. The bulk bands belonging to PCs are plotted in gray, and the interface states are shown by the blue (pseudospin-up), magenta (pseudospin-down) and green curves.}
\end{figure}

Figure~\ref{fig:fig5} shows the band structure for the photonic boundary between different semi-infinite PCs as discussed above. Due to the SHE, two helical interface states are present at the gap, as shown in Figs.~\ref{fig:fig5}(a) and~\ref{fig:fig5}(b). The PC structures are made from the same unit cell but with $\delta$ of opposite sign. There exists a gap between the two interface bands due to the reduced symmetry along the boundary, as illustrated in the inset of Fig.~\ref{fig:fig5}(a). In the panel right to Fig.~\ref{fig:fig5}(a), the phase profile of the interface states at finite wavevector $k_y=\pm0.2\pi/a$ exhibits singularity with a winding phase $2\pi$, indicating finite orbital angular momentum. The phase vortices have opposite winding directions (indicated by yellow arrow) for opposite wavevectors~\cite{Zhang2019Second}. This is featured by the pseudo-spin-momentum locking for helical interface states. %Due to the opposite group velocities of the two %interface states, they propagate along counter-propagating directions. 
While at the boundary of the PCs with $p$-centered unit cell, the symmetry-breaking degree is large enough to widen the gap around $k_y=\pi/a$, as shown in Fig.~\ref{fig:fig5}(b). In this case the system shows SHE as well, but the interface state profile holds two singularities with opposite winding direction [see panels at the right to Fig.~\ref{fig:fig5}(b)].

As mentioned in the above discussions, $\mathrm{PC}_{\delta,0}^{(n)}$ actually locates in the synthetic space of $\mathrm{PC}_{\delta,\eta}^{(m)}$ with $\eta=0.5a$. As such, two interface states are supported at the boundary between $\mathrm{PC}_{\delta,0}^{(m)}$ and $\mathrm{PC}_{\delta,0}^{(n)}$ [see Fig.~\ref{fig:fig5}(c)]. Particularly, a degeneracy appears at $k_y=\pm\pi/a$ due to the glide symmetry at the boundary~\cite{Zhang2019Second}. Similarly, the interface state profile has phase winding at single singularity, defining the pseudospin-up and down distinction. As illustrated in Figs.~\ref{fig:fig4}(a) and~\ref{fig:fig4}(b), $\mathrm{PC}_{\delta,0}^{(m)}$ and $\mathrm{PC}_{\delta,0}^{(p)}$ have different Zak phase, so there is one interface state between the two structures [see Fig.~\ref{fig:fig5}(d)]. Note that the Zak phase is $\theta=\pi$ for $\mathrm{PC}_{\delta,0}^{(m)}$ and $\mathrm{PC}_{\delta,0}^{(n)}$, while $\theta=0$ for $\mathrm{PC}_{\delta,0}^{(p)}$ and $\mathrm{PC}_{\delta,0}^{(q)}$. Therefore, any combination of our proposed PCs guarantees one interface state, which is verified in Appendix \ref{appendix:zak}. 

\begin{figure}[htb]
\includegraphics[width=6.5cm]{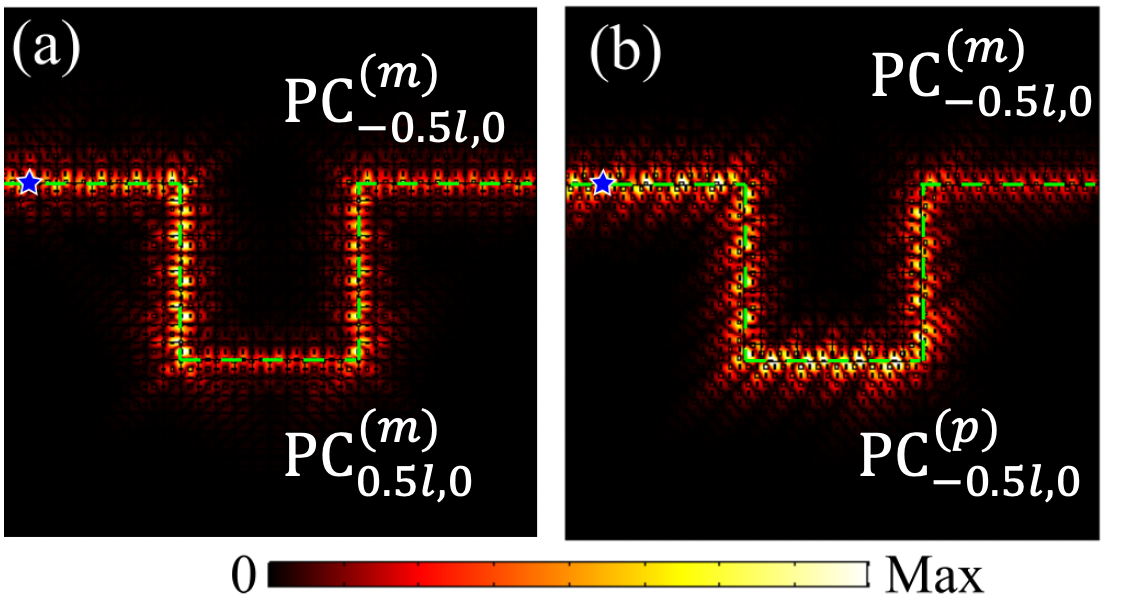}% Here is how to import EPS art
\caption{\label{fig:fig6} Both the pseudo-spin induced interface state in (a) and the Zak induced interface state in (b) can pass through the sharp bends. The green dashed line indicates the interface between (a) $\mathrm{PC}_{-0.5l,0}^{(m)}$ and $\mathrm{PC}_{0.5l,0}^{(m)}$, (b) $\mathrm{PC}_{-0.5l,0}^{(m)}$ and $\mathrm{PC}_{-0.5l,0}^{(p)}$, respectively.  }
\end{figure}

To numerically demonstrate the topologically robust transport of the interface states, we build a U-like interfaces, as shown in Figs.~\ref{fig:fig6}(a) and ~\ref{fig:fig6}(b). A dipolar source (represented by the blue star) with $f=0.494c/a$ is placed at the interface between the PCs with different topological phases. Thus, interface states are excited and propagate along the interface. Clearly, both the pseudo-spin induced interface state in Fig.~\ref{fig:fig6}(a) and the Zak induced interface state in Fig.~\ref{fig:fig6}(b) can pass through the sharp bends.

\section{High-order topological insulator}

Beyond the conventional bulk-boundary correspondence, higher-order topological insulators can be realized in view of the non-zero dipole polarization \cite{Zhang2019Second, Ota2019Photonic}. We start by evaluating dipole polarization $p_{x(y)}$ using the parities ($C_2$ eigenvalues) at high-symmetry $\mathbf{k}$ point of all bands below the gap:
\begin{equation}
\label{eq:eq}
P_{i}=\frac{1}{2}\left(\sum_{n} q_{i}^{n} \text { modulo } 2\right), \quad(-1)^{q_{i}^{n}}=\frac{\eta_{n}\left(X_{i}\right)}{\eta_{n}(\Gamma)}
\end{equation}
where $n$ indicates the index of occupied band, $\mathrm{X}_i$ marks the X or Y point at the first BZ, and $\eta_n(\mathbf{k})$ is the parity of the Bloch mode at momentum $\mathbf{k}$ of the \textit{n}-th band. Accordingly, a dipole topological phase exists when the number of the negative parity of the eigenstate at $\Gamma$ and $\mathrm{X}_i$ is odd. Using Eq.~(\ref{eq:eq}), we identify the $m-$ and $n-$ centered PCs to be topological non-trivial, with 2D dipole polarization $\mathbf{P}=(\frac{1}{2},\frac{1}{2})$. The 2D polarization is connected to 2D Zak phase via $\theta_i=2\pi P_i$ with $i=x,y$ \cite{Liu2017Novel}. As mentioned in the proceeding section, the parity of the inversion operator at X, Y, M cannot be identified for $p-$ and $q-$centered PCs due to their anticommutation relation. Alternatively, the corresponding polarization can be identified to be $(0,0)$ with numerical computation by Eq.~(\ref{eq:eq1}). 

\begin{figure}[b]
\includegraphics[width=8.5cm]{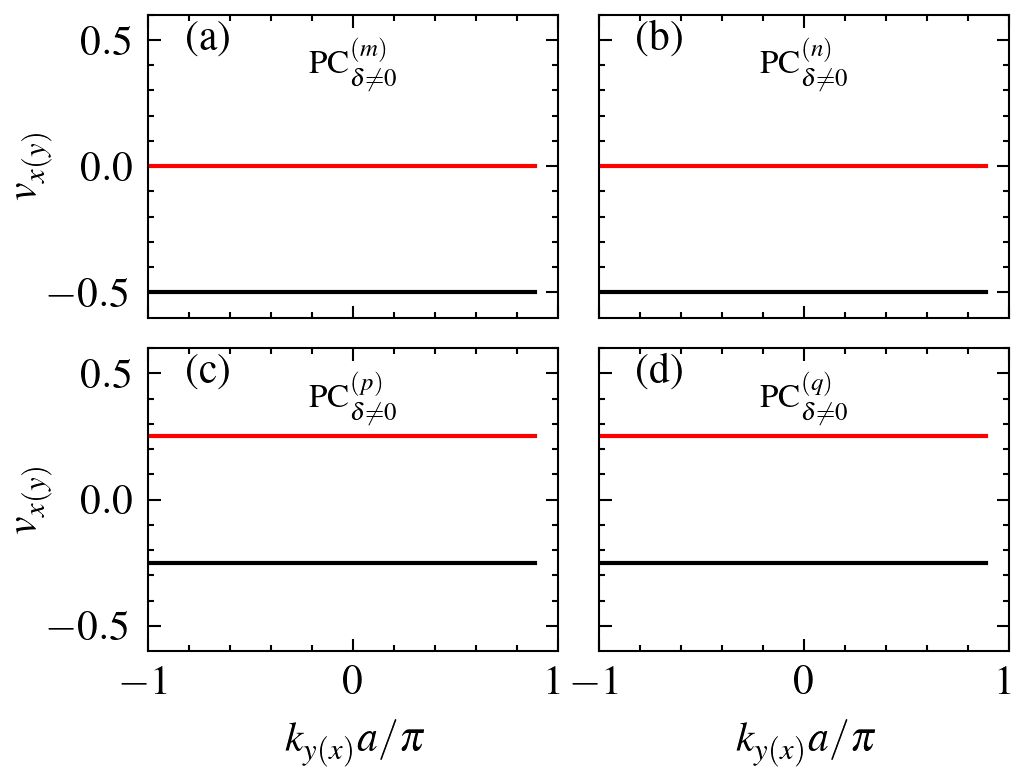}% Here is how to import EPS art
\caption{\label{fig:fig7} Wannier band, $v_{x(y)}$, for the lowest two bands of the proposed PCs. (a)-(d) are the Wannier bands for $\mathrm{PC}_{\delta\ne 0}^{(m)}$, $\mathrm{PC}_{\delta\ne 0}^{(n)}$, $\mathrm{PC}_{\delta\ne 0}^{(p)}$ and $\mathrm{PC}_{\delta\ne 0}^{(q)}$, respectively. }
\end{figure}

To confirm the dipole topology of our PCs, we would explicitly show the dipole moments by the Wilson loop formulation. The Wilson loop operator along the \textit{x} and \textit{y} directions is constructed as $W_{x(y),\mathbf{k}}=F_{x(y),\mathbf{k}+N_{x(y)}\Delta k_{x(y)}}\dots F_{x(y),\mathbf{k}+\Delta k_{x(y)}}F_{x(y),\mathbf{k}}$ with the starting point of the closed loop $\mathbf{k}=(k_x,k_y)$. Here, $[F_{x,\mathbf{k}}]^{m,n}$ is an element of matrix $F_{x,\mathbf{k}}$, an inner product of Bloch eigenstates at adjacent $\mathbf{k}$ point. The eigenvalues of the Wilson loop operator is referred as the Wannier band \cite{Benalcazar2017Electric}. Figures~\ref{fig:fig7}(a)-(d) shows the calculated Wannier bands $v_{x(y)}$ for the lowest two bands of $m,n,p,q$ centered PCs, respectively. The results of Wannier bands in Figs.~\ref{fig:fig7}(a) and \ref{fig:fig7}(b) show that the bulk polarization $P_{x,y}$, which is the integral of the two Wannier bands over the BZ, is $1/2$. The topological distinction between PCs with different dipole polarization guarantees the existence of edge states. Remarkably, those 1D edge states themselves are similar to the 1D photonic SSH model. While for $p$- and $q$-centered PCs, the dipole polarization vanishes, as shown in Figs.~\ref{fig:fig7}(c) and ~\ref{fig:fig7}(d). 

\begin{figure}[t]
\includegraphics[width=8.5cm]{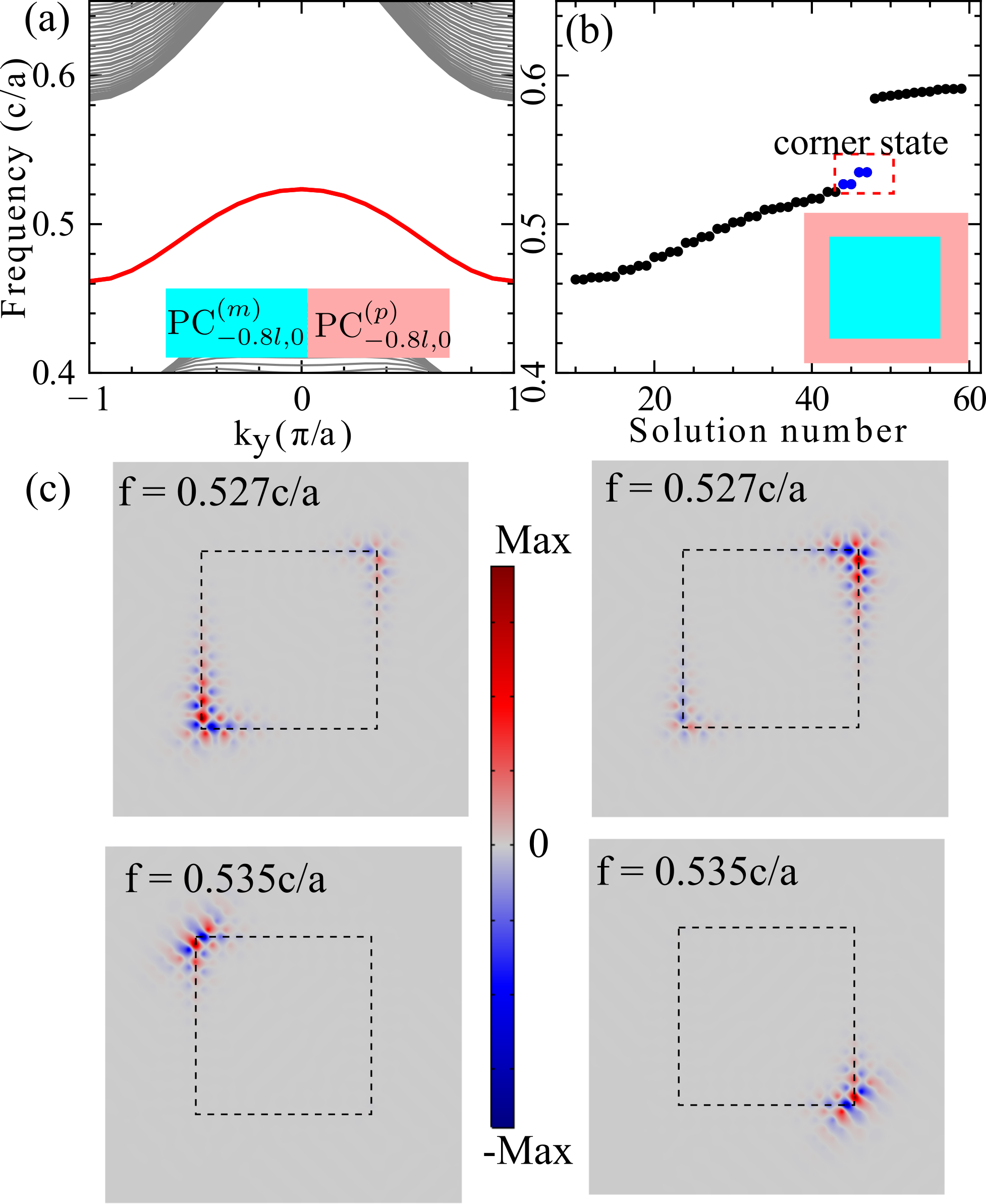}% Here is how to import EPS art
\caption{\label{fig:fig8} (a) Band structure based of the interfacial structures shown in the inset by combination of $\mathrm{PC}_{-0.8l,0}^{(m)}$ and $\mathrm{PC}_{0.8l,0}^{(p)}$. (b) Eigenmodes of the box-shaped structure consisting of $\mathrm{PC}_{-0.8l,0}^{(m)}$ surrounded by $\mathrm{PC}_{-0.8l,0}^{(p)}$. (c) Electric field distribution of the four corner states [blue dots in (b)].}
\end{figure}

According to the above discussion, for $\mathrm{PC}_{\delta,0}^{(m,n)}$ with $\delta\ne 0$, $(P_x,P_y)=(1/2,1/2)$, while for $\mathrm{PC}_{\delta,0}^{(p,q)}$, $(P_x,P_y)=(0,0)$. Figures~\ref{fig:fig8}(a) and \ref{fig:fig8}(b) shows the spectrum of an interface and a box-shaped geometry consisting of $\mathrm{PC}_{-0.8l,0}^{(m)}$ and $\mathrm{PC}_{0.8l,0}^{(p)}$, respectively. As expected from the bulk-edge-corner correspondence elucidated above, four corner states emerge around $f \approx 0.53c/a$. Figure~\ref{fig:fig8}(c) shows that the electric fields are strongly localized at the corners. 

\begin{figure}[t]
\includegraphics[scale=0.9]{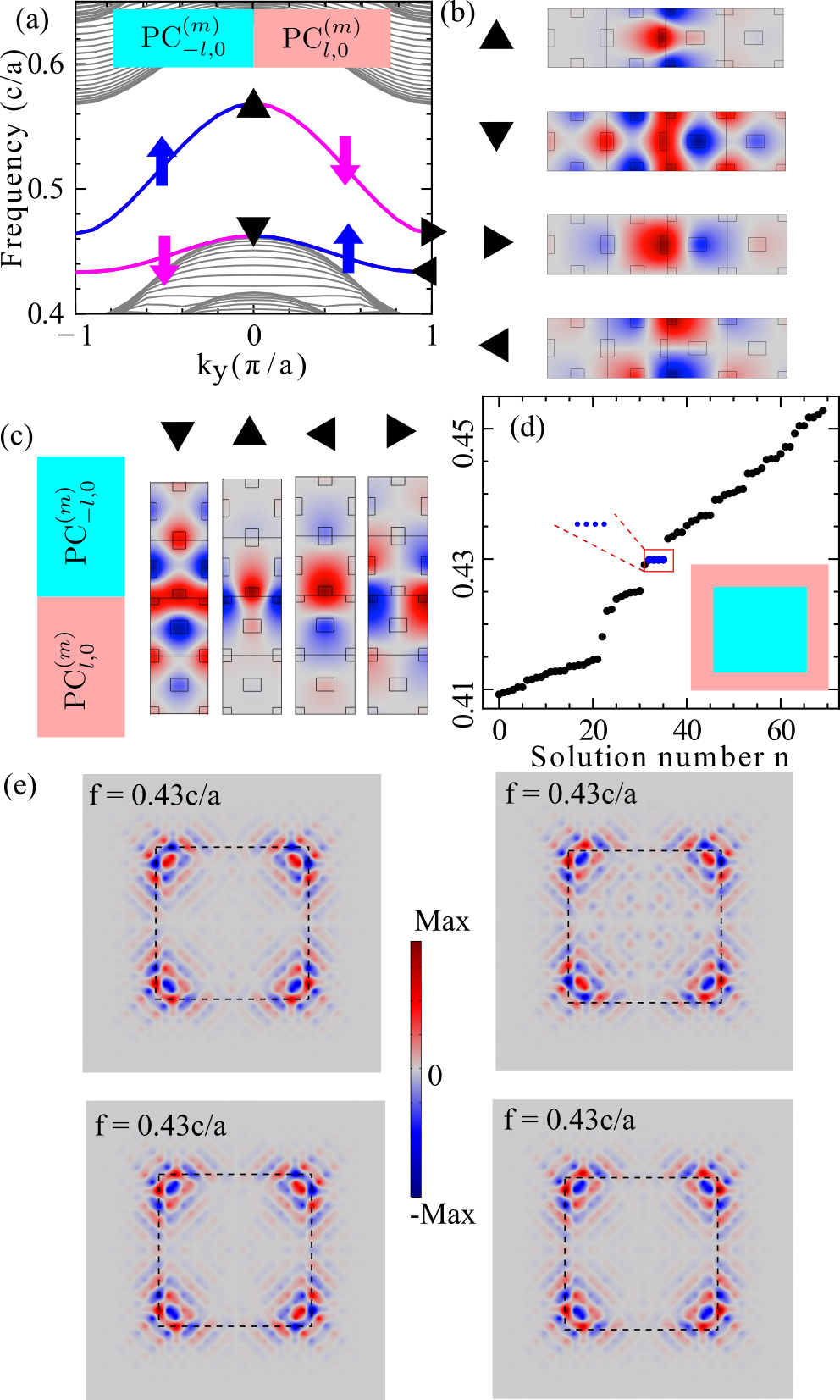}% Here is how to import EPS art
\caption{\label{fig:fig9} (a) Band structure based of the interfacial structures shown in the inset with combination $\mathrm{PC}_{-l,0}^{(m)}$ and $\mathrm{PC}_{l,0}^{(m)}$, respectively. The block sizes are scaled with factor 0.9 and 1.1, in order to form a complete gap. (b) Field distribution Re$E_z$ of the edge mode at the BZ center and edge indicated at (a) for finite cells in \textit{x} directions. (c) Field distribution Re$E_z$ of the edge modes for finite cells in \textit{y} directions. (d) Eigenmodes of the box-shaped structure consisting of $\mathrm{PC}_{-l,0}^{(m)}$ surrounded by $\mathrm{PC}_{l,0}^{(m)}$. (e) Electric field distribution of the four corner states as labeled by the blue dots in (d).}
\end{figure}

We have revealed that SHE is supported in our PCs, and two helical interface states with pseudo-spin up and down, exit at the boundary between \textit{m} and \textit{n} centered PCs with opposite sign of $\delta$. The blue and magenta curves in Fig.~\ref{fig:fig9}(a) represent the corresponding pseudo-spin up and down edge states. In order to obtain a complete gap between the two interface bands due to the reduced symmetry along the boundary, an additional factor $\beta$ is taken to proportional scale the size of the constituent blocks: $w\rightarrow \beta w$, $l\rightarrow \beta l$, $\delta\rightarrow \beta \delta$. Note that the lattice constant \textit{a} remains intact. Here, we choose two specific $\beta=0.9$ and $\beta=1.1$, respectively, acting on $\mathrm{PC}_{-l,0}^{(m)}$ and $\mathrm{PC}_{l,0}^{(m)}$ \cite{Wei2020Realization}. 

The gapped edge states contribute to the broken of glide symmetry at the interface, while the mirror symmetry is preserved which is applied to characterize the parity of the edge states at the BZ center and edge. For both \textit{x} (Fig.~\ref{fig:fig9}(b)) and \textit{y}-axis (Fig.~\ref{fig:fig9}(c)) connection between $\mathrm{PC}_{-l,0}^{(m)}$ and $\mathrm{PC}_{l,0}^{(m)}$, the edge states at BZ center ($k_{x(y)}=0$) are of mirror symmetry. Whereas, two edge state at the BZ edge ($k_{x(y)}=\pi/a$) exhibit mirror symmetry and antisymmetry, respectively. The difference is that the odd (even) edge state at \textit{y} edge has a lower (higher) frequency, while the odd (even) edge state at \textit{x} edge has a higher (lower) frequency \cite{Zhang2019Second}. Therefore, in our PCs, the bulk-edge-corner correspondence is manifested in a hierarchy of dimensions: the topology of both bulk and edge states leads to the corner states, as shown in Fig.~\ref{fig:fig9}(d) and \ref{fig:fig9}(e). Furthermore, we show that the observed topological edge and corner states are also sustained in the corresponding 3D PC slab (see also Appendix \ref{appendix:slab}). Essentially similar results regarding the topological states can be found in PC slabs.

\section{Conclusion}
In summary, we investigate 2D PCs with different topological phase transitions and unconventional higher-order topology. Due to the zone folding of the lattice, the PCs exhibit fourfold degeneracy at Brillouin zone corner and experience topological phase transitions characterized by SHE. Besides, the PCs centered at different points host different Zak phase, SHE, and synthetic Chern number. Furthermore, topological nontrivial polarization gives rise to the topological corner states. The physical origin results from the nonzero dipole polarization. Simultaneously, corner states deriving from multidimensional phase transitions can also be observed, featured by distinct Zak phase of the \textit{x} and \textit{y} edges. Based on the geometrically anisotropic structure, the proposed PCs demonstrates an intriguing way to provide novel designs for topological physics and future related optical applications.

\begin{acknowledgments}
This work was financially supported by Shenzhen Science and Technology Program (No. JCYJ20210324132416040), Guangdong Natural Science Foundation (No. 2022A1515011488), and National R$\&$D Program of China (No.2018YFB1305500).
\end{acknowledgments}

\appendix
\section{Band folding induced fourfold degeneracy}
\label{appendix:fold}
\begin{figure}[b]
    \centering
    \includegraphics[width=8.5cm]{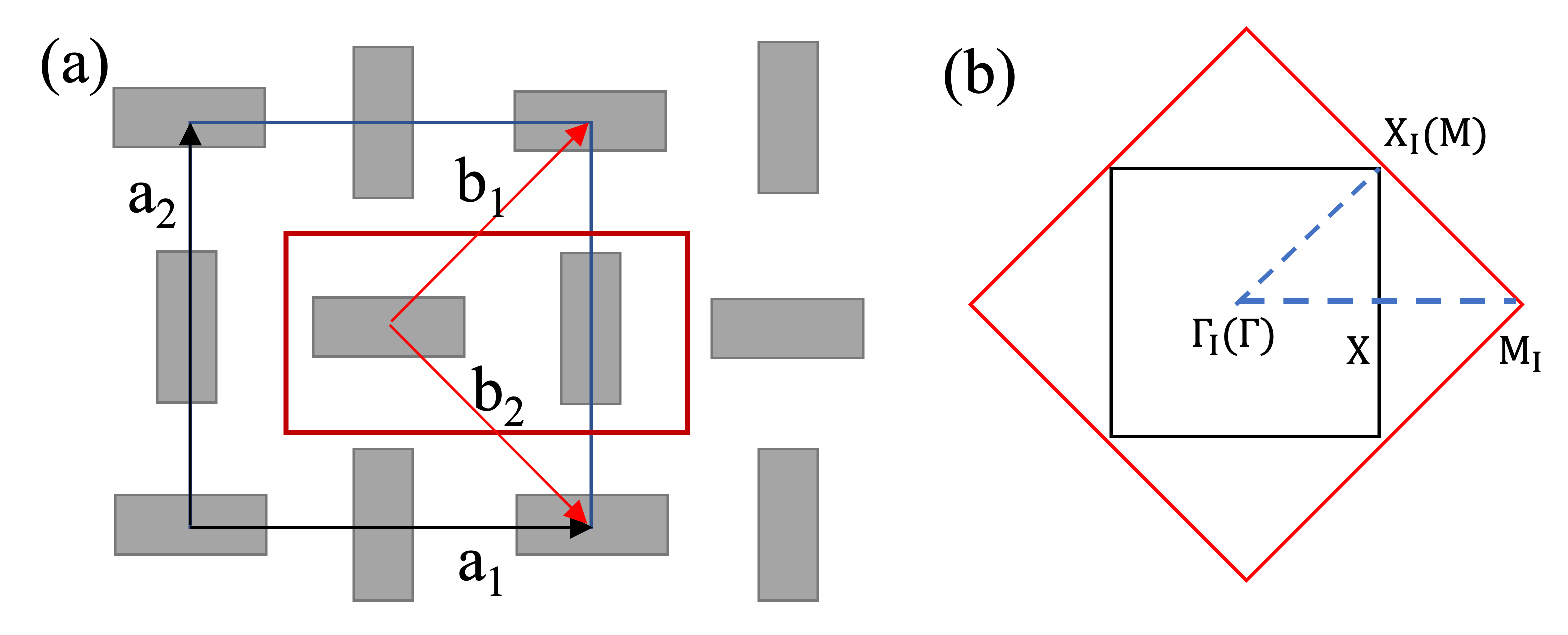}
    \caption{Schematic of the band folding effect. (a) The configuration of the PCs with primitive (red box) and enlarged unit cell (black box), and (b) the corresponding BZ.}
    \label{fig:fig10}
\end{figure}

2D PCs consist of rectangle meta-blocks whose long axis orientates alternatively toward x and y directions, as shown in Fig.~\ref{fig:fig10}(a). The primitive cell is outlined by the red box, whose lattice vectors are, respectively, $\mathbf{b}_1$ and $\mathbf{b}_2$. The enlarged unit cell outlined by the black square have two lattice vectors, $\mathbf{a}_1$ and $\mathbf{a}_2$. Clearly, the primitive unit cell holds the glide symmetry $G_x:(x,y)\rightarrow (-x,y+a/2)$. Then, in combination with the time-reversal symmetry T, $(G_xT)^2:(x,y)\rightarrow (x, y+a)$ can be obtained. Note that the translation along the \textit{y}-axis can be taken as $\mathbf{b}_1-\mathbf{b}_2$. As such, $(G_xT)^2=e^{i(\mathbf{k}_1+\mathbf{k}_2)\cdot(\mathbf{b}_1-\mathbf{b}_2)}$ with $\mathbf{k}_1$ and $\mathbf{k}_2$ the reciprocal lattice vectors of the primitive lattice. When $k_1-k_2=\frac{\pm\pi}{a/\sqrt{2}}$, $(G_xT)^2=-1$ satisfies the Kramer’s double degeneracy. As such, double degeneracy is present along the line $X_1-M_1$ as shown in Fig.~\ref{fig:fig10}(b). Similarly, double degeneracy is present at the whole BZ edges in combination with the glide symmetry $G_y:(x,y)\rightarrow (x+a/2,-y+a/2)$. 

When the unit cell is enlarged, the glide symmetries $G_x$ and $G_y$ are preserved. In this case, $(G_xT)^2=e^{i(\mathbf{k}_x+\mathbf{k}_y)\cdot \mathbf{a}_2}$ is satisfied. Here, $\mathbf{k}_x$ and $\mathbf{k}_y$ are the reciprocal lattice vectors of the enlarged unit cell. Then, $(G_xT)^2=-1$ is satisfied when $k_y=\pm \pi/a$. Similarly, the glide symmetry $G_y$ yields the Kramer’s double degeneracy at the line $k_x=\pm \pi/a$ . Then the BZ edges of the enlarged unit cell are double degeneracy. On the other hand, the enlargement of the unit cell gives rise to the band folding along the BZ edges of the enlarged unit. Then fourfold degeneracy occurs at momentum M labeled in Fig.~\ref{fig:fig10}(b).

\section{Band diagram of PCs featured by Zak phase}
\label{appendix:zak}
\begin{figure}[htb]
    \centering
    \includegraphics[width=8.5cm]{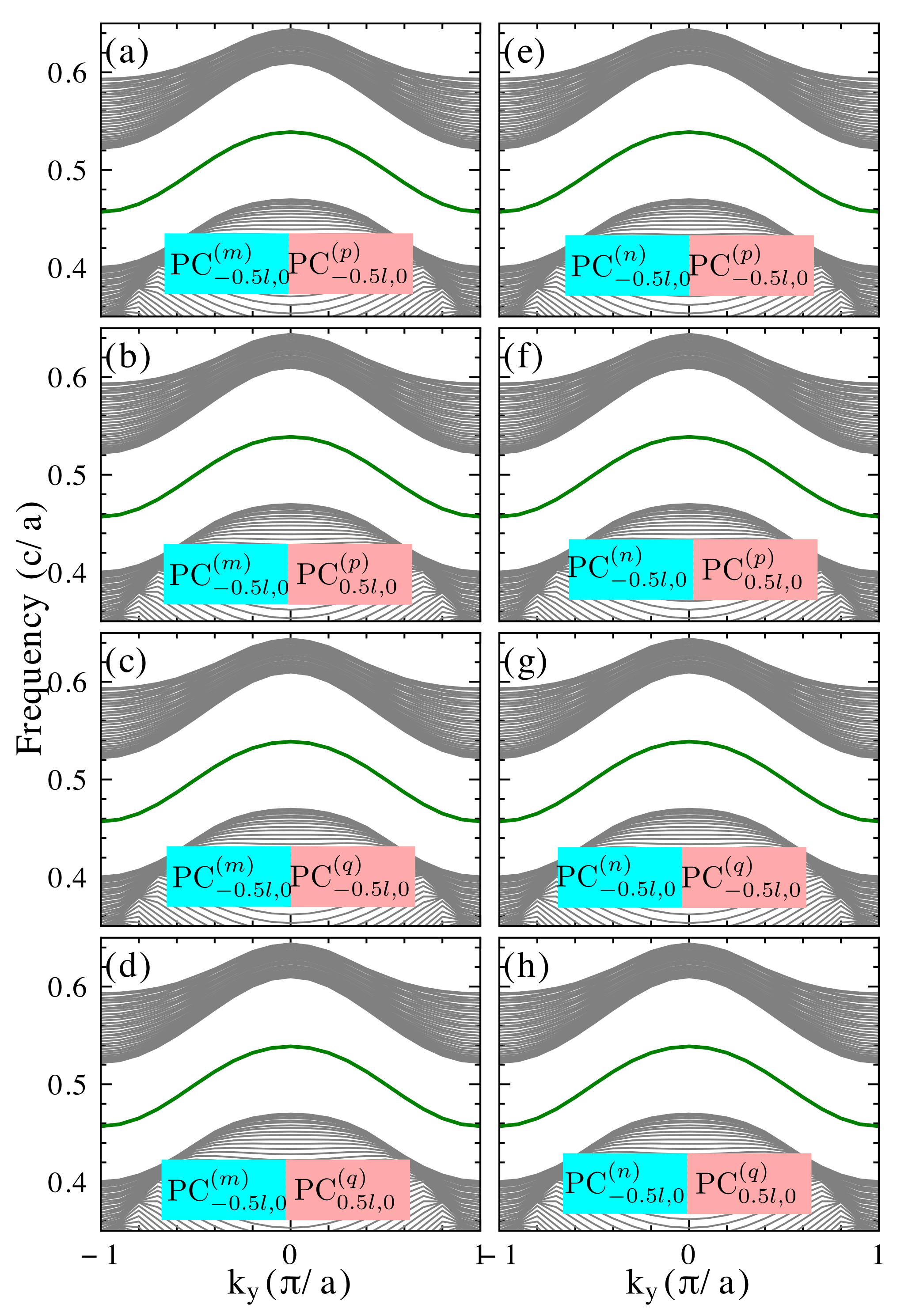}
    \caption{Projected band diagram for the interfacial structures shown in the inset. The left-side PC in the left panel is the \textit{m}-centered unit cell with $\delta=-0.5l$, and the right-side PC in the right panel is the \textit{n}-centered unit cell with $\delta=-0.5l$.}
    \label{fig:fig11}
\end{figure}

In the main text, Fig.~\ref{fig:fig5}(d) shows the band diagram of the combination between $\mathrm{PC}_{-0.5l,0}^{(m)}$ and $\mathrm{PC}_{-0.5l,0}^{(p)}$ with different Zak phase. The other results of the unique combination between the \textit{m/n}- centered PCs and \textit{p/q}- centered PCs are shown in Fig.~\ref{fig:fig11}. This is determined by their distinct Zak phase, as revealed in Fig.~\ref{fig:fig7} 

\section{Interface states by synthetic Chern number}
\label{appendix:chern}
\begin{figure}[htb]
    \centering
    \includegraphics[width=6.5cm]{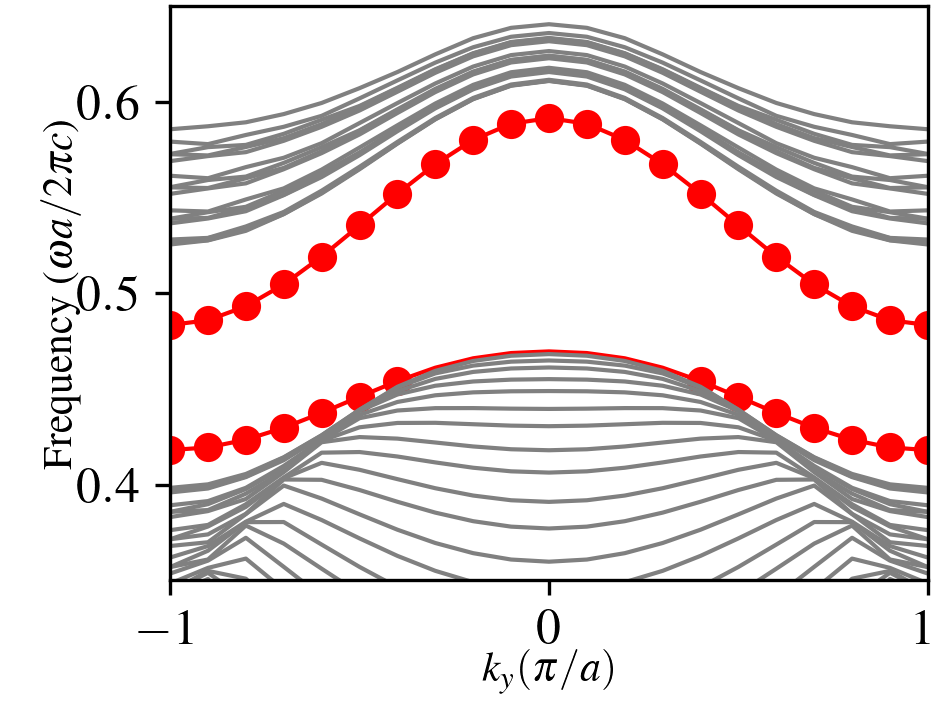}
    \caption{Dispersion bands of interface states for the interfacial structures $\mathrm{PC}_{-0.5l,0}^{(p)}$ and $\mathrm{PC}_{-0.5l,0.4a}^{(p)}$}
    \label{fig:fig12}
\end{figure}

As illustrated in Fig.~\ref{fig:fig4}, the Zak phase winding in synthetic space yields the synthetic Chern number 2. The result of this non-zero Chern number is the presence of the interface states for the interfacial structure with translational displacement. Figure~\ref{fig:fig12} shows the dispersion for the interfacial structures $\mathrm{PC}_{-0.5l,0}^{(p)}$ and $\mathrm{PC}_{-0.5l,0.4a}^{(p)}$. Clearly, two interface states emerge at the band gap, which is consistent with the synthetic Chern number 2.

\section{Interface states by pseudospin-2 and Zak phase in complementary PC slabs}
\label{appendix:slab}
\begin{figure}[htb]
    \centering
    \includegraphics[width=6.5cm]{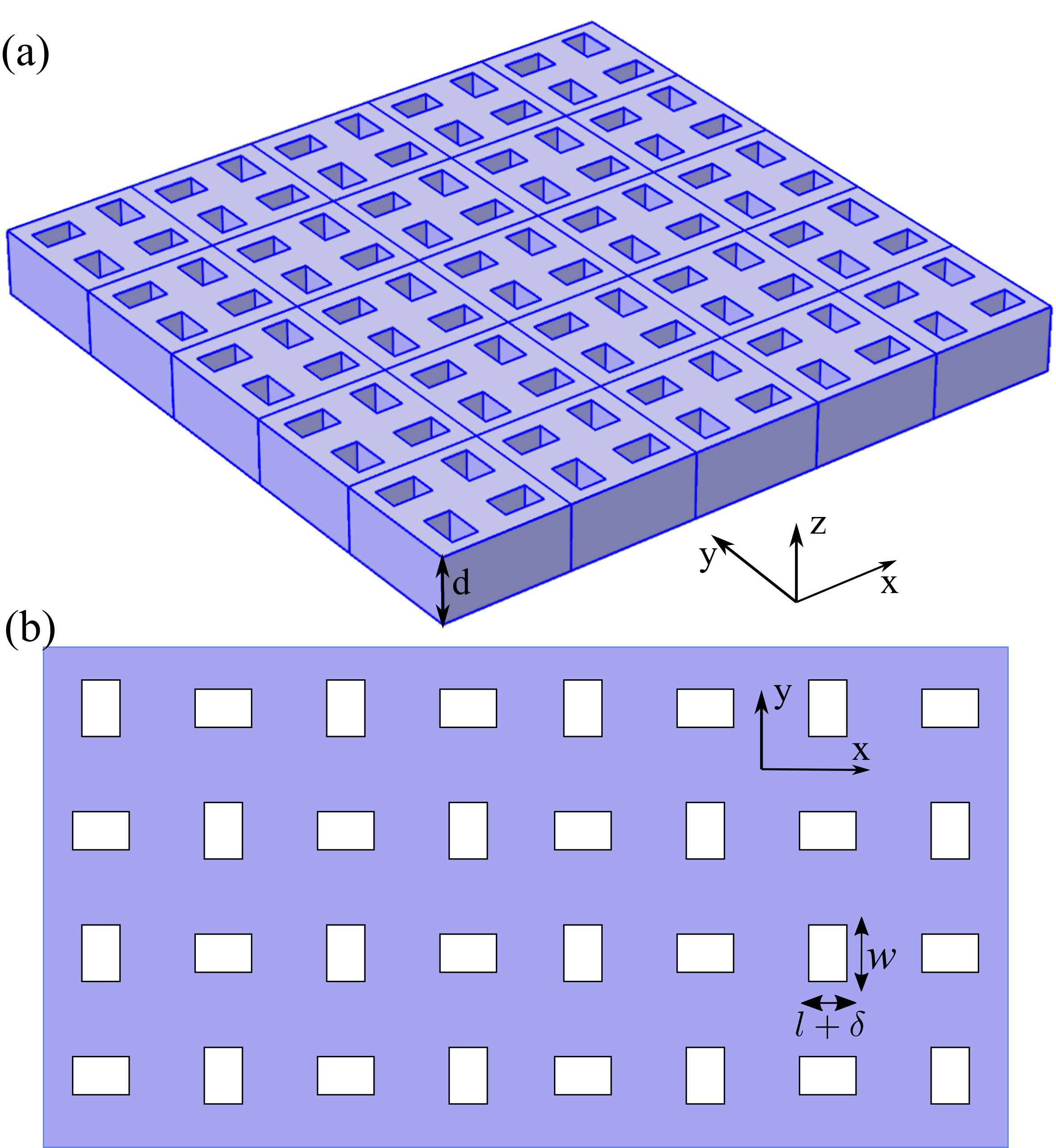}
    \caption{(a) Schematic of the PC slab. (b) Top view of (a).}
    \label{fig:fig13}
\end{figure}

For a more interesting generalization to 3D system, we consider the corresponding PC slab [see Fig.~\ref{fig:fig13}(a)]. The PC slab is composed of air holes drilled in a dielectric slab (e.g., 3D version complementary structure to the PC studied), and the geometry symmetry features of the holes is similar to the rectangular rods in Fig.~\ref{fig:fig2}. The dimensions of the holes are $w=0.33a$, $l=0.3a$, and the thickness of the slab is $d=0.5a$ (see Fig.~\ref{fig:fig13}(b)). 

\begin{figure}[htb]
    \centering
    \includegraphics[width=8cm]{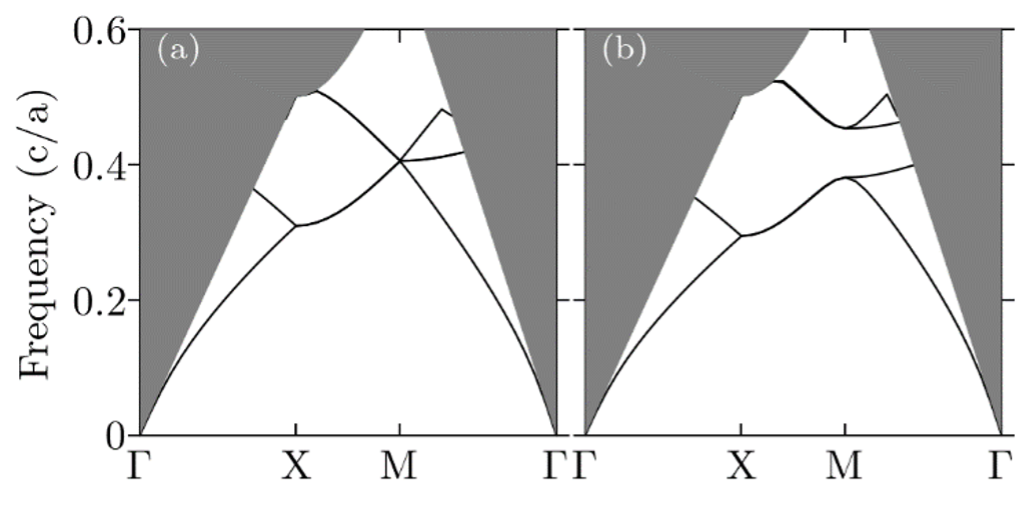}
    \caption{Photonic band structure of the TE-like mode for (a) $\delta=0$ and (b) $\delta=\pm 0.5l$. The gray region represents the light cone.}
    \label{fig:fig14}
\end{figure}

Figure~\ref{fig:fig14} shows the band structure of the transverse electric (TE)-like mode. Certainly, the introduction of $\delta$ would split the fourfold degeneracy at ‘M’ [see Fig.~\ref{fig:fig14}(a)] into two pairs of double degeneracy, as shown in Fig.~\ref{fig:fig14}(b). This is in accordance with the two-dimensional case, e.g., Figs.~\ref{fig:fig2}(b)-\ref{fig:fig2}(d).

\begin{figure}[htb]
    \centering
    \includegraphics[width=8.5cm]{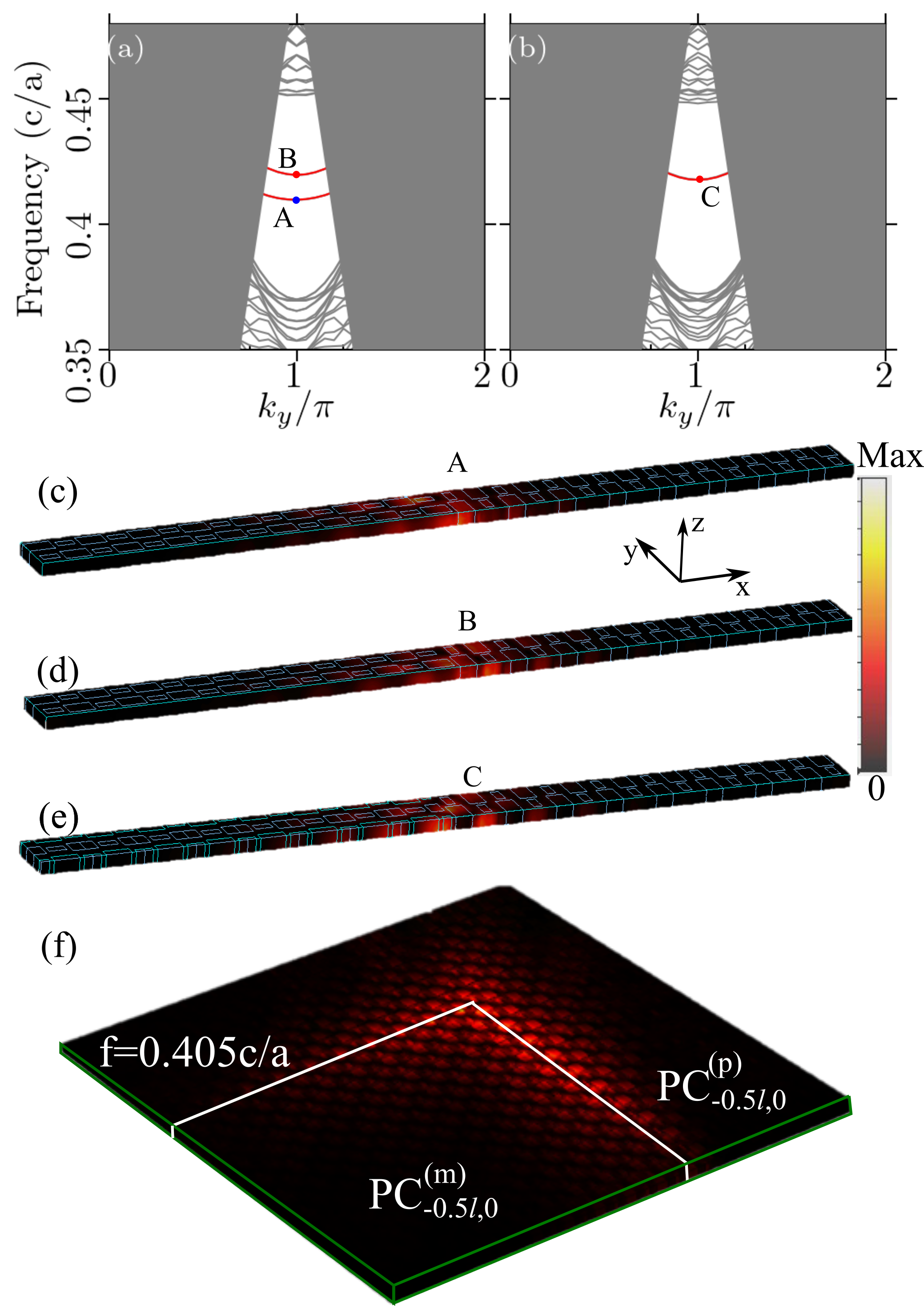}
    \caption{Band structure of the interfacial structures with combination (a) $\mathrm{PC}_{-0.5l,0}^{(p)}$ and $\mathrm{PC}_{0.5l,0}^{(p)}$ and (b) $\mathrm{PC}_{-0.5l,0}^{(p)}$ and $\mathrm{PC}_{-0.5l,0}^{(m)}$ made by the PC slab in Fig.~\ref{fig:fig13}. (c-e) The field distribution $\vert \mathbf{E}\vert$ of the interface states labeled in (a) A and B and (b) C, respectively. (f) The field distribution $\vert \mathbf{E}\vert$  of the corner state at $f=0.405c/a$ for $\mathrm{PC}_{-0.5l,0}^{(p)}$ and $\mathrm{PC}_{-0.5l,0}^{(m)}$. The gray region in (a) and (b) marks the light cone. Red curves denote the bounded interface states inside the bandgap.}
    \label{fig:fig15}
\end{figure}

For the PC slab, the topological phase transition can also be observed and gives rise to the photonic spin Hall effect. Figure~\ref{fig:fig15}(a) shows the projected band structure with the combination of $\mathrm{PC}_{-0.5l,0}^{(p)}$ and $\mathrm{PC}_{0.5l,0}^{(p)}$ PC slab shown in Fig.~\ref{fig:fig13}. Obviously, two interface states are present at the bandgap, whose field distributions $\vert \mathbf{E}\vert$ are shown in Figs.~\ref{fig:fig15}(c) and \ref{fig:fig15}(d), respectively. Figure~\ref{fig:fig15}(b) shows that only one interface state is present at the bandgap of the interfacial structures made by the $\mathrm{PC}_{-0.5l,0}^{(p)}$ and $\mathrm{PC}_{-0.5l,0}^{(m)}$ type PC slab. This is due to the different Zak phase of these combined unit cells. The field distribution of this interface is shown in Fig.~\ref{fig:fig15}(e). All these phenomena for PC slab are consistent with the 2D counterpart, which is essentially determined by the spatial symmetry of the studied lattice. 

% \bibliography{references}% Produces the bibliography via BibTeX.
%

\end{document}